# Assistant, Parrot, or Colonizing Loudspeaker? ChatGPT Metaphors for Developing Critical AI Literacies

Anuj, Gupta; [0000-0002-5336-927X](#) ; Yasser, Atef; [0000-0002-2755-8800](#); Anna Mills; [0009-0001-5094-1461](#); Maha Bali; [0000-0002-8142-7262](#)

## ABSTRACT

This study explores how discussing metaphors for AI can help build awareness of the frames that shape our understanding of AI systems, particularly large language models (LLMs) like ChatGPT. Given the pressing need to teach "critical AI literacy", discussion of metaphor provides an opportunity for inquiry and dialogue with space for nuance, playfulness, and critique. Using a collaborative autoethnographic methodology, we analyzed metaphors from a range of sources, and reflected on them individually according to seven questions, then met and discussed our interpretations. We then analyzed how our reflections contributed to the three kinds of literacies delineated in Selber's multiliteracies framework: functional, critical and rhetorical. These allowed us to analyze questions of ethics, equity, and accessibility in relation to AI. We explored each metaphor along the dimension of whether or not it was promoting anthropomorphizing, and to what extent such metaphors imply that AI is sentient. Our findings highlight the role of metaphor reflection in fostering a nuanced understanding of AI, suggesting that our collaborative autoethnographic approach as well as the heuristic model of plotting AI metaphors on dimensions of anthropomorphism and multiliteracies, might be useful for educators and researchers in the pursuit of advancing critical AI literacy.

## INTRODUCTION

In May 2023, several senior undergraduate students at the Texas A&M University were denied graduation even though they had all the requisite credits to get a degree. The reason? One of their teachers had failed them, on the charge that they used ChatGPT, a text generation AI chatbot that has taken the world by storm, to plagiarize their class assignments. His methodology to check his students' work for AI related plagiarism involved him putting the students' writing into ChatGPT itself and asking it whether it had written them. The chatbot erroneously claimed that it had, something it does even for famous passages from classic literature like *Crime and Punishment* (Klee, 2023). The teacher took the chatbot's response at face value, assuming, it seems, that this statistical model was a coherent entity that had access to knowledge of its previous actions. His action hinged on the way he imagined ChatGPT, and his imaginings had practical consequences for students. As it turned out, imagining ChatGPT as a being with knowledge of its past leads to wrong conclusions about its capacities. In fact, it has no access to past outputs and has no algorithm to assess whether text was machine-generated or not. This incident is one among many where the everyday use of AI algorithms within digital writing practices—something that has been likened to an oil-spill within an ecosystem (Bender, 2023), coupled with certain conceptualizations of what they are, is causing immense harm. Such incidents highlight fears of an impending "textpocalypse" (Kirschenbaum, 2023), or a situation when machine generated automated text, "synethic text devoid of human agency or intent" (Kirschenbaum, para 3), overwhelms the existing infrastructures of our text based societies.

To prevent such harm, there have been many calls to cultivate "critical AI literacy" (Bali, 2023). This call, which enters into conversation with existing approaches to critical digital literacy (Raffaghelli et al., 2020, Zawacki-Rickter et al., 2019; Selber, 2004), has inspired a wide range of teachers to develop different aspects of critical AI literacies with open educational practices as a guiding framework (Mills et al., 2023). For example, Anna Mills has meticulously compiled an open-source database of sources related to AI and writing in higher education (Mills, 2022). Raffaghelli et al (2020) call for raising awareness and questioning of agendas in technological and data-driven solutions in education and this can be extended to AI use of any kind in education.  As mentioned earlier, Zawacki-Rickter et al (2019) call for more critical reflection on pedagogical and ethical implications as well as risks of using AI in higher education.

One of the ways in which we've found it productive to enact that approach to education in our discussions on Twitter (note: partway through writing this paper, Twitter's name changed to X), is through analyzing the public discourse around ChatGPT, especially the kind of metaphors that are present in conversations around such algorithms. In this paper, our main objective is to demonstrate that approach by unpacking the public discourse around AI in general and LLM products like ChatGPT in particular, by focusing on the role that metaphors play as key discursive phenomena that frame our thinking and consequently our behavior. By doing this, we present "metaphor reflection" as an approach through which we can develop metacognitive awareness about how our attitudes and behaviors towards AI are shaped by ideological forces in the public spheres which we inhabit. This process of metaphor reflection can be a resource for all of us as educators, scholars, and students interested in enacting critical AI literacies (Bali, 2023) in playful, critical and nuanced ways in our respective contexts.

In doing so, we are guided by two research questions:

1) How are metaphors in public discourse on AI, particularly LLMs such as ChatGPT, shaping our understanding and behavior towards these systems?
2) What implications, ethical considerations, equity and accessibility concerns, research directions, and attitudes do these metaphors enable/mediate/engender?

To answer these questions, in this paper, we begin by introducing the role that metaphors play in shaping public discourse, after which we review existing research on critical AI literacies, on the role of metaphors in education, and how these two domains can be synthesized. We then describe our methodology and our positionalities to make explicit how our identities and experiences influence our interpretation and research. Specifically, we describe how we, the authors (located and coming from the US, India and Egypt, an undergraduate student, a PhD candidate/graduate teaching associate, and two professors, one of whom is a writing teacher and one is mainly a faculty developer), followed a methodology of digital collaborative authoethnography. We first generated a diverse list of metaphors about AI and LLMs, drawing from peer-reviewed literature, popular discourse, blogs, social media feeds, workshops, and conversations with people outside the education sphere, with the aim of capturing a representative range of metaphors that we had been exposed to, and then we qualitatively coded these metaphors to unpack how they reflect, contradict, and shape our conceptualizations of AI. In this paper, we present the findings and analysis that arose from our exploration/contemplation, which highlight the value of metaphor as a stimulus for understanding AI in playful, critical and nuanced ways that are dynamic. We end with implications for teachers and researchers and possible directions for integrating the power of metaphors into critical AI literacy.

**LITERATURE**

CRITICAL AI LITERACY AS A GOAL FOR ARTIFICIAL INTELLIGENCE APPLICATION IN HIGHER EDUCATION

An important means of helping students engage with AI tools in higher education settings that we see emerging around us is "Critical AI Literacy" (Bali, 2023). In her definition, Bali emphasizes the nature of criticality in two ways. She writes, "One is critical as in critical thinking, as in skepticism and questioning. The other is critical as in critical pedagogy, so focusing on social justice dimensions and inequalities something may exacerbate, reproduce or create. Another is critical, as in, critique it for its potential harms, and critique the credibility/accuracy of its outputs/outcomes (this can also come from a combo of the other two criticals anyway)". When applied to the teaching and use of AI tools, this critical attitude includes "helping students recognizing the limitations and potential of AI – what they gain or potentially lose when using it," while keeping in mind that "there isn't one right way, there is the way that works for you within the parameters of your situation" (Bali, 2023).

This perspective is concomitant with findings from researchers like Zawacki-Rickter et al. (2019) and Raffaghelli et al. (2020) who in the past have highlighted a predominance of instrumental or functional approaches to AI in higher education, both in research and in pedagogical practice. For example, Zawacki-Rickter et al. (2019) found in their multi-country, multi-disciplinary systematic review of research on AI in education that researchers predominantly spoke about functional or instrumental literacies in relation to the adoption of AI in four key areas: "profiling and prediction, assessment and evaluation, adaptive systems and personalisation, and intelligent tutoring systems" (p.1) with a "dramatic lack of critical reflection of the pedagogical and ethical implications as well as risks of implementing AI applications in higher education" (p.21). Zawacki-Rickter et al. (2019) had foregrounded an "almost lack of critical reflection of challenges and risks of AIEd, the weak connection to theoretical pedagogical perspectives, and the need for further exploration of ethical and educational approaches in the application of AIEd in higher education" (p.1). Similarly, Raffaghelli et al.(2020) have noted that in spite of growing research on how algorithms and data have inbuilt racial and gender-based biases, systematic reviews of educators' data literacies have pointed out a predominance of instrumental, technical and data management perspectives and lack of critical or social perspectives in data literacies.

To ground our understanding of critical AI literacies, we find it productive to contextualize it within Selber's (2004) "multiliteracies" framework. This framework is a macro-level heuristic that helps guide teachers and students to develop core digital literacies in three domains: functional, critical, and rhetorical. Through "functional literacy", we can think of computers (and AI by extension) as tools, with students as effective users of technology with the objective of developing effective employment opportunities. Through "critical literacy" on the other hand, we can conceptualize computers and AI as "critical artifacts", with students as questioners of technology who engage in informed critique. Finally, through "rhetorical literacy" Selber (2004) nudges us to think of computers (and AI by extension) as hyptertextual media that students engage with as producers of technology to create reflective praxis. While the first two categories are often easy to understand for most teachers, the third one called "rhetorical literacy" is often difficult to grasp given the pejorative meanings that the term "rhetoric" often gets associated with in common parlance. What Selber (2004) means by "rhetorical literacy" is a combination of functional and critical literacies that helps students to use or build technologies in a critically informed manner. For example, "rhetorical activities like Web design demand both effective computer use and informed critique" (Selber, 2004, p.25). The key to remember here is that these categories are not water-tight, but often overlap and build on each other.

Integrating Selber's (2004) multiliteracies framework for computer literacies with the findings of Zawacki-Rickter et al. (2019) and Raffaghelli et al. (2020), we now see critical AI literacies (Bali, 2023) as a heuristic that helps teachers develop curricula along three dimensions: functional, critical, and rhetorical in ways that do not prioritize any single dimension but rather help students learn how to merge them. We believe that such a critical AI literacy framework can help augment existing research and practise around incorporation of AI tools in higher education settings. For example, researchers like Wang et al. (2023) show that creating a supportive environment is essential to enabling students' learning about AI. In their work, supportive environments consist of "facilitating conditions" like accessibility to technology as well as "supportive social norms" (the extent to which students perceive other social members' desire to want them to learn about AI).

Our work helps extend this conversation on creating supportive environments for AI learning by arguing that supportive learning environments around AI should also make space for critical exploration of the nature and ethics of AI. Students will have a range of emotions about AI as well as foundational questions about AI. Literacy practices must support students' sense of agency as they explore these questions. Literacy practices should not dictate outright implementation or rejection of these tools but rather encourage and facilitate thoughtful and student-centric decision making. Such critical AI literacy practices can also help tackle a major barrier that Renz and Hilbig (2020) had found to implementation of AI tools: a "lack of data understanding and insufficient data sovereignty" (p.17) in potential users. Critical engagement with AI can include discussion of how many contemporary digital AI tools use and produce data. Such discussion can encourage students to informed decisions about how and whether to use particular AI systems based on how those systems may use their data.

Our work is thus grounded within a critical AI literacy framework that seeks to support students in agentively engaging with these technologies. We seek to present one of the ways in which such literacy can be operationalized in classroom contexts.

IMPACT OF METAPHORS USED ON COGNITION, BEHAVIOR, AND EDUCATION

While in common parlance, we often think of metaphors simply as ornamental devices or rhetorical flourishes in our writing, within psychology, there is a strong tradition of research in psychology, linguistics, and philosophy that shows how metaphors in everyday discourse play an important role in shaping our thinking and behavior. Most famously, Lakoff and Johnson (2003) have put forward their "conceptual metaphor theory." They have argued that the conceptual systems that structure "what we perceive, how we get around in the world, and how we relate to other people" are "fundamentally metaphorical in nature" (p.4). In terms of method, they use linguistic evidence "to identify in detail just what the metaphors are that structure how we perceive, how we think, and what we do" (p.4). This approach resonates with work like that of the philosopher Schön (1993) who presented a generative metaphor theory which foregrounds the building of new metaphors as "a process by which new perspectives on the world come into existence" through what he called the "meta-pherein" or "carrying over of frames or perspectives from one domain of experience to another" (p.137). An example from Lakoff and Johnson (2003) helps illustrate various facets of both conceptual metaphor and generative metaphor theories. They ask us to think of the ways in which we speak about arguments. When we say things like "She won the argument", or "they defend their theory against his attacks", what we essentially ascribe to is a conceptual metaphor of arguments-as-war. This makes us think of arguments as a battle that needs to be won, something that consequently shapes our behaviour towards arguments in confrontational ways and often leads to a degeneration of public discourse on social media comment sections. One of the ways in which we attempt to change this social reality then, is through generating and disseminating new metaphors that help us conceptualize arguments in very different ways and consequently argue differently. For example, Lakoff and Johnson (2003) suggest that if we were to conceptualize arguments-as-dance, then perhaps we would value the production of harmony, rhythm and beauty while arguing, something that might reduce the toxicity we often see on our social media feeds.

These theories have had a deep impact on various academic and applied fields like public policy, political campaigns, advertising and education. Within education, Lukeš (2019) argues that there are three broad ways in which teachers and researchers have applied these theories. These include using metaphors as "an invitation to enter", as an "instrument to grasp knowledge with", and as a "catalyst to transform understanding."

As a form of invitation, "teachers like to reach for metaphors relying on the familiar. This gives the learner a chance to grasp onto something while they build up sufficient mental representations of the new domain". This can help provide initial emotional support towards understanding but does not fully create the understanding itself. For example, he shows that when teachers help students understand electric current as a flow of water, they can make them feel less anxious about visualizing what the flow of electrons looks like. However, this can lead to partial understandings because students might start to equivocate water with electricity and wonder how water gets inside electric wires. To tackle this, Lukeš (2019) argues that teachers can push students to use metaphors in the second way, i.e. as an instrument to grapple with concepts. This involves thinking through the similarities as well as the differences between the two domains of knowledge that a metaphor engenders, as well as thinking about multiple metaphors that can help illuminate different aspects of the target concept. Going back to the water metaphor, Lukeš (2019) writes that while the water metaphor helps students understand the flowing properties of electricity, it does not elaborate on how electrons rub against each other to produce energy. For this latter property a different metaphor, something like small balls rubbing against each other, can be more helpful. These first two modes involve teachers using existing metaphors to help aid students' understanding in progressively complex ways, and are closely aligned with Lakoff and Johnson's (2003) "conceptual metaphor" idea. The third form of use that Lukeš (2019) advocates for, i.e. metaphor as catalyst, involves building on the initial forms of understanding to create new metaphors to produce new understandings, something that is in sync with Schön's (1993) "generative metaphor" idea. Lukeš elaborates, "here the metaphor becomes a process without an end. It spurs new mixtures and remixtures as one finds out more about the two (and often more) domains. Unlike with instrumental and invitational metaphors, it is no longer important that the metaphor be apt. It is just important that it is useful for new understandings or the possibilities of these new understandings." In terms of the example of electricity as water and as ball-like particle metaphors, this stage would involve students creating new metaphors that approximate these complex layered meanings and also realizing that no metaphor can single-handedly capture all aspects of a target concept.

In terms of research applications, we see many ways in which such layered understandings of metaphors' impact on cognition and behaviour in education are operationalized. Tham et al. (2021) have studied the metaphors that students use to conceptualize new digital technologies and digital literacies and demonstrated how teachers can harness this knowledge to scaffold their digital literacy acquisition. In a different context, Martins (1991) showed how "waking up sleeping metaphors" in biology textbooks, like "the picture of egg and sperm drawn in popular as well as scientific accounts of reproductive biology" which "relies on stereotypes central to our cultural definitions of male and female" (p. 486), rather than the most recent scientific research, helps us in "becoming aware of their implications" and reduces "their power to naturalize our social conventions" (p. 501).

Our work is thus inspired by this entire body of teacher-scholars who have used metaphors as a tool to scaffold instruction and support students to question and transform their understanding of key concepts.

METAPHORS FOR AI IN PUBLIC DISCOURSE, EDUCATION AND BEYOND

Since the launch of OpenAI's *ChatGPT,* terms like "AI", "LLMs", "Generative AI" etc. have been circulating in public discourse, especially social media discourse, with increased frequency and intensity. Through our own experiences on social media, we have seen how a lot of this discourse is often marked by a lot of hype, usually in the form of either existential fears or in the form of techno-utopic optimism. We've also seen scholars like Bender (2022) making rigorous efforts to cut through such hype and Johnson (2023) urging teachers that "the discourse of crisis needs to give way to more generative thinking" (p. 173).

There are some researchers who've been exploring the role of metaphors in shaping public understandings in this discourse on AI in general and ChatGPT in particular. With regards to the role of metaphors for AI, Khadpe et al. (2020) used an experimental approach to present an AI conversational agent to users where they used metaphors describing their agent that varied along axes of "warmth" and "competence" and then surveyed their participants about their intentions to use the agent based on these descriptions. Interestingly they found "the intention to adopt and desire to cooperate decreases as the competence of the AI system metaphor increases" and "users are more likely to co-operate and interact longer with agents portraying high warmth, but we do not observe significant impact of warmth on users' intention to adopt" (p. 15). More recently, Andersen (2023) explored three very popular metaphors that are being used to conceptualize ChatGPT, that of artificial intelligence (which we often don't notice is a metaphor itself!), tool and collaborator. By analyzing media sources and critiquing their framing of these metaphors, Anderen (2023) goes on to provide medical or surgical metaphors for ChatGPT's engagement with human writing instead as a means to encourage more critical engagement with it: "AI-generated compositions may even resemble blood products (with aggregated contributions from vast numbers of people), stem cells (which derive from particular people but which may be used to generate something new), or donor organs and tissues (which enable the transplant of whole structures from one person to another)" (p.8). Through these generative metaphors, they recommend that teachers should be thoughtful about the kind of metaphors they use in their assignments as they talk about ChatGPT or similar technologies, and also encourage students to notice how different technical documents around them, like university policies and academic guidelines, employ metaphors for AI.

To sum up, we see our research as expanding Andersen's (2023) approach to thinking about metaphors that we use for ChatGPT and AI technologies in the classroom as a means to developing critical digital literacies. We have built on their approach by bringing it into conversation with the larger body of research on the role of AI metaphors in tech adoption like that of Khadpe et al. (2020); application of metaphor theory in education (Tham et al. 2021; Lukeš, 2019; Martins, 1991; ); and calls to develop critical AI literacy (Bali, 2023) that counter-balance the more instrumental and functional approaches to AI literacy (Raffaghelli et al., 2020; Zawacki-Rickter et al., 2019). Through this synthesis we have developed a practical and accessible way in which teachers can help develop students' critical AI literacy by helping them explore existing AI metaphors as well as creating new ones in a playful, critical and nuanced manner.

# POSITIONALITY

As this is a collaborative autoethnography, we share below our positionality so we can be transparent with the reader as to how our identities and experiences may influence and shape our subjective responses to AI and its metaphors.

*Author 1:* I am a PhD candidate and Graduate Associate at the University of Arizona in the US in the discipline of Rhetoric and Composition. Before this, I helped build one of India's first college-level writing programs at the Young India Fellowship at Ashoka University. Through the last ten years as a researcher, teacher, and student of writing, I have been deeply interested in how Web 3.0 technologies like social media, AR/VR and AI shape our writing, thinking, and behaviors. While traditionally trained as a literary scholar and then as a social scientist, over the last few years, I have been taking classes in Python and Machine Learning to help engage with the impact of AI and LLMs on education and writing. In my dissertation now, I want to continue this work at the intersection of the Humanities and Technology, to help students and teachers engage in more critical and agentic ways with Web 3.0 phenomena like LLMs.

*Author 2:* As a student, I have the opportunity to observe the informal and potentially hidden applications of ChatGPT in academic contexts through collegial discussions and class/ group collaborations. This allows me to gain insights that may not be readily accessible to professors and academia as a whole, primarily because of the existing boundaries between students and their mentors. While this novel and groundbreaking technology poses many challenges at hand, students and educators still hold varying perceptions about ChatGPT and its counterparts. To provide a better understanding of my own background, I am an undergraduate student, having honed my skills in different areas. As opposed to my geographical location in Egypt and being an Egyptian individual, disabled, and young, I have traversed diverse levels of education and come from an emerging economy. I also approach technology differently, having to adhere to the available accessibility guidelines and screen reading applications' adaptability to these artificial intelligence tools. As a consequence, I try to shun the staggering digital inaccessibility impulses and advocate for an equitable and inclusive community. Therefore, metaphors, as a relatively understood communication method amongst educators, had been found one of the best methods to establish a common ground around AI and to provide a comprehensive understanding around its use, ethical considerations, and potential risks. To me, this had an everlasting impact on my work as a student, as to why we should use AI in the first place, when to use it, and the controversy about whether, or not, to count it as a new facilitative tech.

In my position as a student, however, I also take into account the ease of taking shortcuts. As detrimental as it may sound, students would not think about personal growth, unfair advantages, or reputation while committing this misconduct. Students are often pressured to secure grades, believing that they are the only measure of success, and with less attention drawn towards the risks such as tarnishing one's academic record. In such circumstances, the allure of shortcuts triggers students towards using this Large Language Model while sacrificing the opportunities for lifelong learning, overshadowing the importance of integrity. As a result, I realize that, in the quest of expediency, students will more likely lack meaningful skills, including critical thinking, research, and the ability to synthesize thoughtful analysis.

*Author 3:* I am a community college writing teacher, but I grew up in Silicon Valley; my father was a software engineer who worked on machine learning systems we would now call AI, like handwriting recognition and spelling correction. I shared his fascination with philosophical questions about consciousness, the line between the mechanical and the human, the ways in which language can be mechanized, and the absurdities that can result. When I realized what language models could do in June 2022, I was riveted and felt like I was circling back to familiar questions. I wanted to figure out not just what these models meant on a practical level for the teaching of writing but on a deeper level how it would be best to think about them and describe them to students.

As I encountered critiques of AI discourse like Bender et al.'s (2021) Stochastic Parrots paper, I could see how much was at stake in the metaphors we use for AI and also how useful they could be. I began to explain language models to my students, I found myself reaching for metaphors and pausing to consider the way each metaphor might shape our classroom discourse about AI. As the author of an Open Educational Resources textbook about writing and the English Discipline Lead for the Open Educational Resources Initiative of the Academic Senate of the California Community Colleges, I have been immersed in OER discussions that often include open pedagogy and ways to collaborate with students to create educational materials. Thus I have gravitated toward exploring ways in which academics and students could engage in similar and intersecting practices to shape our critical understanding of AI and our decisions about whether or how to use it.

*Author 4:* My main role right now is as a faculty developer in my institution, to support other educators in their teaching. So when ChatGPT came on the scene, I was one of the main people figuring out how to support the institution and professors to figure out how to respond. Why was I the main person in my department doing this work?

My undergraduate degree was in computer science, and my graduation project was a neural network, so I understand how machine learning works, even if the last time I created one was 20 years ago. I later studied eLearning and did a PhD in education, and digital education and digital literacies are among my areas of expertise. I've been teaching an undergraduate course on digital literacies for several years now, and we always tackled ethical issues and biases in AI, long before ChatGPT came on the scene.

So this semester, in parallel, I integrated more discussion of AI in my class, and my colleagues in my department and I did a lot of testing and research on AI in education, and we offered multiple community conversations, workshops, and panels on the topic of AI in education. Some of my earliest testing of ChatGPT was in January outside the main semester, and so I did a lot of testing at home with my 11 year old child beside me - I was heavily influenced by her own reactions to it, her initial fascination with it, then her quick disillusionment with it.

In my public scholarship, I often ask people about metaphors of AI on Twitter and during keynotes and workshops: and I feel like their answers said a lot about their attitudes towards it.

This idea of using metaphors came from a conversation with my mom and my cousin. My mom is the one who gave the "fast food" metaphor, which I always use as an example and ask people to unpack what that metaphor implies. The metaphor of "cake making" (Bali, 2023) as one way to decide whether using AI is a good or bad idea came from a conversation with my cousin who works in HR and was concerned about skills of graduates and readiness for the workplace.

I love how we can use biomimicry to understand the world. Machine learning is itself already a kind of metaphor and neural network a kind of biomimicry, right? I tend to use metaphors in my teaching and blogging very often, so doing so for AI made sense. I also noticed recently that the Quran uses metaphor and simile very frequently, especially relating to nature (so biomimicry again).

**METHODOLOGY**

The methodology used was a digital collaborative autoethnography. Autoethnography "seeks to describe and systematically analyze personal experience in order to understand cultural experience" (Ellis, et al., 2011). Collaborative Autoethnography (CAE) aims to provide a rich, multi-dimensional perspective on a topic as it builds on the experience of the diverse authors. The CAE process usually begins with each individual contributing their personal reflections/narratives, which we then collectively discuss, revisit, analyze, and relate to the literature (Geist-Martin, et al., 2010).

Autoethnographic research "challenges the hegemony of objectivity or the artificial distancing of self from one's research subjects" (Chang, et al., 2013, p. 18). Each of us came into our experiences with AI and metaphors with our own positionalities which affected our perspectives and biases towards AI and the kinds of metaphors that stood out for us. CAE falls under the range of interpretive/critical research traditions, as a participatory research methodology which does not adhere to scientific/positivistic measures of validity and rigor. Since "all understanding is already interpretation" and "the interpreter is always already part of what is being interpreted" (Nixon, 2012, p. 33), we emphasize also that "all understanding necessarily involves an element of self-understanding" (p. 34, citing Gadamer, 2001).

Our particular CAE was conducted digitally, as we live in different parts of the world, and also collected our metaphors via interactions online: via Twitter/X, during Zoom calls with each other, and during virtual workshops. We also had our own discussions virtually via email, Twitter/X, Google docs, and Zoom calls. We conducted our analysis on a Google Sheet, with different sheets for each of us to analyze each metaphor. CAE as a social science methodology that centers human experience diverges from much of the literature around AI, which, according to Zawacki-Richter et al's (2019) systematic review, has been largely first-authored by STEM and computer science authors (p.20) who rarely, if ever, use such a methodology. When we think about how AI may influence teaching and life beyond university, the perspectives of learners and educators can add value as end users of the technology that computer scientists immersed in the technology itself may not be able to contribute. Between the four authors here, the reader is getting the perspective of an undergraduate student, a PhD student, two people who teach writing, and an educational developer who also teaches digital literacies. The reader is also getting both a US and Egyptian perspective. Our goal here is not to reach an objective conclusion as to what metaphor is best for AI or how the metaphor influences all people, but rather to convey a nuanced interpretation of a variety of metaphors and how they may influence people in different positions differently.

The use of metaphor introduces another layer of subjectivity in our interpretation. We were also inspired by the literature on metaphors that helped us understand how they shape cognition and behavior across domains especially in education (Lukeš 2019; Lakoff and Johnson, 2001; Schön, 1993). Using this body of work, we realized that methodologically, metaphors can serve as a potent interface between the subjective experiences and impressions we have about phenomena and the collective, socio-cultural and ideological forces that shape them. We do not claim the generalizability of our findings, but instead highlight different angles, a process Laurel Richardson (1997) calls crystallization.

OUR PROCESS OF DATA COLLECTION

In order to come up with the list of metaphors, we had discussions via email, in Google Docs and their margins, Zoom conversations, and on Twitter threads. We also discussed how we would analyze the metaphors, which aspects of them to highlight. Some of the metaphors we use here have been used in peer-reviewed literature (e.g. "stochastic parrots", Bender et al, 2021), in more popular literature and discourse (e.g. "calculator for words", Wilson, 2023), or on someone's blog (e.g. "good people", Mollick, 2023) or Twitter feed (e.g. "cake making", Bali, 2023). We also included metaphors for AI we discovered from participants in workshops we gave (Author 1 and Author 4 explicitly solicit these) and in conversations with others outside the education sphere (e.g. "fast food"). In our discussions, we sometimes came up with new metaphors to highlight certain aspects of ChatGPT that other metaphors did not capture (e.g. "oil guzzler") or that might have cultural relevance to us personally (e.g. "Aage kuan piche khai" in Hindi, which literally means "a well in front, a ditch behind", similar to the English expression"between a rock and a hard place" and "fahlawi" in Arabic, a term that has a negative connotation of confident liar or bullshitter). We also built on each others' metaphors in order to give the metaphor more complexity (e.g. "Venus fly trap as one type of flower"). Our goal in this data collection strategy was to cast a wide net to capture metaphors about AI and LLMs circulating in our public spheres, both popular as well as more obscure ones. As must be evident by our processes of collection and producing metaphors, we were inspired by both conceptual and generative approaches to metaphors that we have read about in the literature on the role of metaphors in cognition, behavior and education (Lukeš 2019; Lakoff and Johnson, 2003; Schön, 1993). For further reference, please see our dataset which has been archived on Zenodo: https://zenodo.org/records/10468056

DATA ANALYSIS

In order to analyze the metaphors collaboratively, we created a master Google sheet with the list of metaphors and their source, then each individually had a separate sheet to reflect on with several questions corresponding to each of the metaphors, including

1. What feelings does the metaphor evoke?
2. Where does it help us understand AI?
3. What kind of perspective does it push the listener to imagine about AI?
4. Where is the metaphor helpful and harmful?
5. Is it a generally positive or negative view of AI?
6. In what ways does the metaphor showcase accessibility or equity issues related to AI?
7. How does the metaphor portray the relationship between humans and AI?

The next step was to synthesize all of our reflections on each metaphor, and to find a way to thematize the various metaphors and how they portray and influence our perceptions, attitudes and knowledge about AI. For example, some metaphors make us suspicious of AI (e.g. "wolf in sheep's clothing") while others tend towards making us appreciate it (e.g. "assistant" or "life saver") and still others tend to make us fear it (e.g. "terminator"). Some metaphors use nature (plants or animals) and others refer to human analogies. The synthesis occurred on Google docs with each of our names beside our reflections, then on Google docs and in discussion over Zoom, we selected key words and quotes from our own reflections to keep, and included some elements of our own live dialogue around each metaphor. Some unplanned dialogue between the four of us occurred on public Twitter and was integrated as well.

Our overall approach towards this synthesis was driven with multiple, intersecting motivations. On the one hand, we were striving towards thematic coherence in ways that would allow us to see the larger patterns of meaning-making about ChatGPT that have been shaping our collective thinking about it, but on other hand, we were also mindful to keep the thematic patterns we were generating flexible enough to allow for each of our individual subjectivities to be reflected in them, while also holding space for the generation of new metaphors for both us and our readers. For further reference, please see our analysis of our dataset which has been archived on Zenodo: https://zenodo.org/records/10468056

**FINDINGS**

We present the results of our analysis in two forms below. First, in Fig 1, we present a diagram that plots the metaphors we collected and generated about ChatGPT along two thematic dimensions that emerged through our analysis process. We call the first dimension, what you see on the X-axis of the diagram as the "anthropomorphic dimension", which on the left extreme includes metaphors that we coded as conceptualizing ChatGPT and AI to be completely human like, and on the right extreme includes non-human metaphors. In the middle of this axis are metaphors that visualize ChatGPT in animalistic or plant-like terms. We used the second dimension, which you see on the Y-axis of the diagram, to represent Selber's multiliteracies framework as discussed above, from functional to critical to rhetorical approaches. We've plotted all the metaphors that we studied together based on how they relate to these two dimensions, in Fig 1.

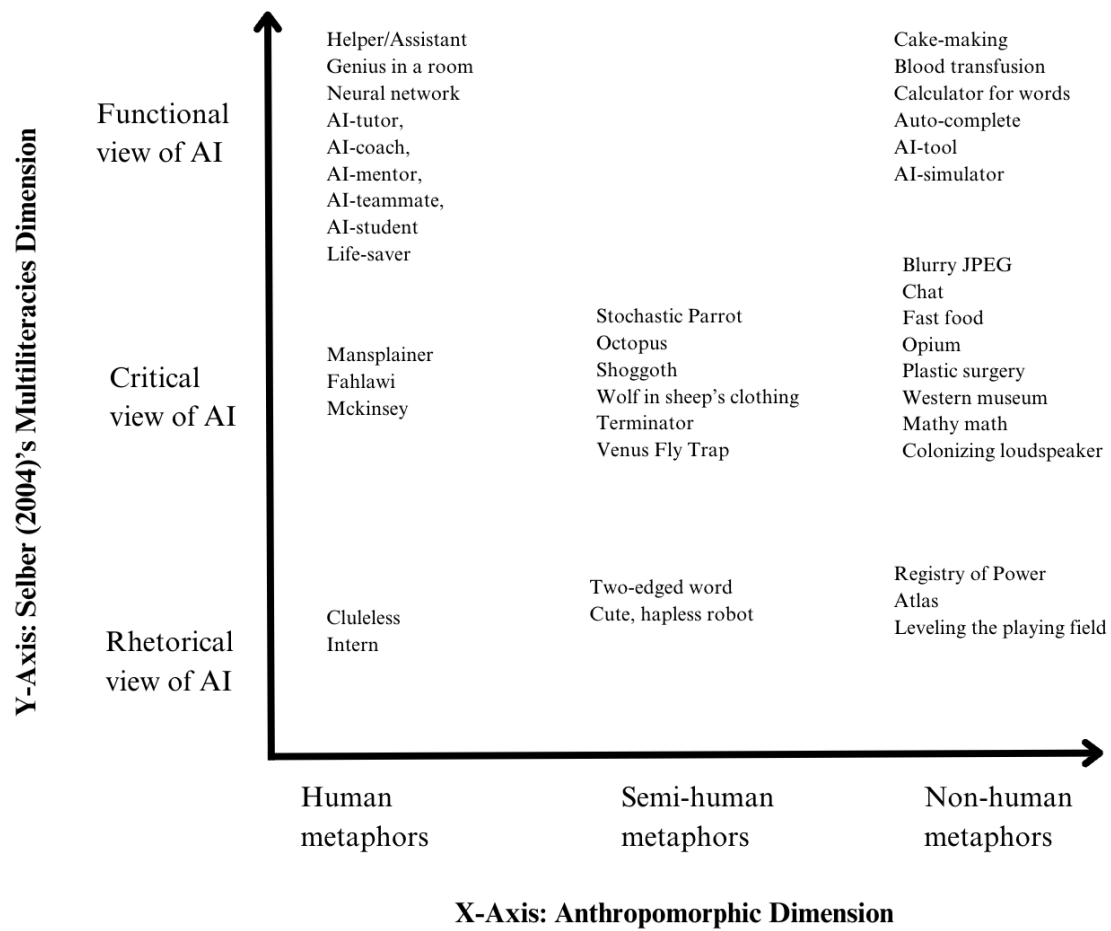

Fig 1: Analysis Results in Diagram Format

Second, to ensure accessibility for the wider community of readers particularly for those with visual disabilities and other learning disabilities, including dyslexia and dyspraxia, we have curated our results from Fig 1, and also present our results in Table 1 in a grid format. In this table, the columns represent the "anthropomorphic dimension", while the rows represent the "evaluation dimension" that we discussed above.

Table 1: Analysis Results in Grid Format

| Metaphor | Human | Animal/Plant/Humanoid? | Nonhuman/inanimate |
|---|---|---|---|
| **Functional view of AI** | Helper/Assistant<br>Genius in a room<br>Neural network<br>AI-tutor, AI-coach, | | Cake-making<br>Blood transfusion<br>Calculator for words<br>Auto-complete |

|  | AI-mentor,<br>AI-teammate,<br>AI-student<br>Life-saver |  | AI-tool<br>AI-simulator |
| --- | --- | --- | --- |
| **Critical view of AI** | Mansplainer<br>Fahlawi<br>Mckinsey | Stochastic Parrot<br>Octopus<br>Shoggoth<br>Wolf in sheep's clothing<br>Terminator<br>Venus Fly Trap | Blurry JPEG<br>Chat<br>Fast food<br>Opium<br>Plastic surgery<br>Western museum<br>Mathy math<br>Colonizing loudspeaker |
| **Rhetorical view of AI (Blends functional and critical)** | Clueless intern | Two-edged word<br>Cute, hapless robot | Registry of Power<br>Atlas<br>Leveling the playing field |

## DISCUSSION

Our analytical and reflective process of meaning-making and thematizing was influenced by our initial guiding questions, which emphasized plotting and tabulating our metaphors around analyzing the feelings and understandings each metaphor evokes, the ways it harms and helps, the attitude it promotes, and so on. We had several discussions that led to the emergence of many themes about our understanding of how metaphors about ChatGPT and AI have been shaping our thoughts and actions, and how the use of new metaphors could potentially catalyze new attitudes and behaviors towards them. These had focused on how these metaphors influence our feelings and our thinking about the potential impact of LLMs and AI in our lives. We describe these themes below.

THEME 1: IS AI SENTIENT? THE ONTOLOGICAL NATURE OF LLMS

The first major theme that emerged in our discussions related to the x-axis in our graph or the "anthropomorphic dimension". We noticed that a majority of the metaphors like "helper", "assistant", "student", "coach", "mentor", "team-mate", "neural network", "mainsplainer", "fahlawi", "clueless intern", "shoggot" etc. conceptualized LLMs as human or human-like implying that it is sentient in the way that humans are. These metaphors pushed us to describe or define what we mean when we say sentience or consciousness and then reflect on whether or not

LLMs meet those criteria. Our discussion revolved around criteria like: creativity in language production, ability to reason, ability to make errors. As we probed further, another theme emerged in our discussion - even if it appears like LLMs have many criteria for sentience, are they only mimicking what they see humans do at a surface-level or do they have a deeper understanding and mechanisms through which these surface-level sentience features are produced? To some of us, the fact that at a deeper level LLMs use statistical methods to produce these surface-level features, made them fundamentally different from how humans produce these features, which is why LLMs felt like an *imitation of sentience rather than sentience itself*. This is why metaphors that imply a lower level sentience or no sentence at all, like "stochastic parrots", "octopus", "venus fly trap", "a flood", "blurry JPEG", "registry of power", "calculator" etc., felt more appropriate. To others, however, since we couldn't precisely define the deep structures in humans that produce these surface-level features, nor could we prove a complete lack of any level of statistical computation, it was difficult to say whether or not LLMs are sentient, and therefore using only non-sentient metaphors might obscure or at least deemphasize some of the existential questions that LLMs bring. For example, we know that what LLMs do is much more complex than what a calculator does, and much less predictable. Describing it as such can be dangerous as it oversimplifies its functionality as well as its impact in the world, especially its potential harm.

At this point, we also came across Gallagher (2023)'s recent interview study of machine learning engineers where he recounts a popular joke that many of his participants told him: "Machine learning is written in Python. AI is written in PowerPoint" (p. 150). What this implies is that while ML or machine learning is a specific, computational technique used by computer scientists to automate tasks, AI often has marketing connotations that has a discursive currency in the grant writing and fundraising world, not necessarily in the scientific world. This alerted us to the fact that choosing whether to use sentient or non-sentient metaphors while describing AI is not simply a theoretical discussion, but is rather sometimes a strategic rhetorical strategy used by AI developers to shape market perceptions. This helped us transition to our next theme, which shifted the conversation from ontology to ethics or pragmatics where we started probing the effect that using certain kinds of metaphors have on different kinds of users.

THEME 2: MORAL/ETHICAL DIMENSION OF LLM METAPHORS

In theme 2, we reflected on the Y-axis or evaluative dimension of our heuristic where we looked at how different metaphors push us to adopt functional, critical or rhetorical views of LLMs. Through reflecting on the set of metaphors we collected, we realized that many LLM products like ChatGPT or Bard are often presented with human-like features to create a human-computer interaction experience of "chatting" that establishes comfort in its user base and to facilitate interaction by helping activate mental models of interaction that users might be familiar with. Google's Bard, for example, calls itself our "creative and helpful collaborator". We noticed how these metaphors which are most popular and present in the marketing of user interface of many of these chatbots present LLM models as not just human-like but particular kinds of servile and harmless human entities that are there to help and augment humanity without threatening the agency of the humans or "controlling" them. These are "functional" metaphors because they facilitate functional interactions. However, if only these metaphors are used, then they potentially obscure the actual mechanistic processes that produce these products and also prevents us from seeing any harm that they might do. This is why other metaphors that are more critical and functional are also important. In these categories we started seeing how scholars like Bender *et al.* (2021) have created counter metaphors like that of "stochastic parrots", seeking to help humans see the non-human, mechanical, statistical nature of the technology that powers these products and alerting them to their potential dangers. This metaphor helped us realize that anthropomorphizing a "stochastic parrot" involves fooling people into thinking they are speaking to someone with human-like agency, intelligence, emotional investment, integrity and responsibility which can lead to astonishing disastrous effects.

Another kind of critical metaphor pointed toward the ways in which AI systems and language models in particular reproduce and amplify existing power dynamics. Metaphors like a "western museum" and a "colonizing loudspeaker," for example, arose in workshops Author 4 facilitated at the American University of Cairo and for Equity Unbound, and in conversations amongst the authors of this article (inspired by the work of Owusu-Ansah, 2023). "Colonizing loudspeaker" refers to the way in which the prevalence of colonial perspectives in the training dataset for ChatGPT and other language models will tend to make its output reflect colonial viewpoints and language norms. Widespread adoptions of these models will spread these perspectives rapidly while creating the impression that they are neutral since they come from an automated system.

As we probed through other metaphors, we started to realize that only focusing on critical metaphors that point us towards the harms that ChatGPT-like technologies create, might prevent us from recognizing the benefits that these technologies might bring to marginalized users - representing "critical" as in social justice oriented (Bali, 2023) and interrogating potential harm (Raffagheli et al, 2020). For example, in one of Author 1's classes, a multilingual student had pointed out that using ChatGPT made him feel on par with his native English-speaking white peers. By giving him confidence to articulate his ideas in a language of power, Author 1 felt that ChatGPT was "leveling the playing field" for him. While Author 1 appreciated that, he was however careful to cite his use of ChatGPT but the student felt confused how to do it in classes where professors had banned it. Author 1's working with this student is like what Selber (2004) calls a "rhetorical literacy" that tries to blend functional and critical understandings of a technology to develop a reflective praxis. Similarly, in Author 4's class of non-native speakers of English learning in English, students mentioned how AI can help students read something that is too complex for them to fully understand on their own, by summarizing it for them and answering their questions about it. As we discussed the potential benefits and harms of functional, critical and rhetorical metaphors, we started transitioning to the final theme in our discussions, where we thought about how these kinds of discussions can contribute to developing critical AI literacy.

THEME 3: METAPHORS AS A PLAYGROUND FOR DEVELOPING CRITICAL AI LITERACIES IN FUNCTIONAL, CRITICAL, AND RHETORICAL MANNERS

One thing that stood out was the open-ended and generative nature of the format we chose; "play" and "exploration" might serve to characterize the ethos of our exercise, including the spreadsheet plus live Zoom discussion plus social media engagement methods we used. The fact that we kept a fairly loose structure around the spreadsheet and deadlines allowed us to add in metaphors as we saw them arise in social media, our workshops, and blogs. It allowed us to explore metaphor in relation to ChatGPT from a standpoint of curiosity, humor, pleasure, inquiry and critical concern for existing and potential harms. We included emotional responses, and we did not feel pressured to come up with "right" answers or to debate and correct each other's answers, in contrast to sometimes more polarized social media and other public discussions of these questions.

The ethos of "play" and "exploration" gives us hope that dialogue around metaphors might play a role in classroom development of an exploratory critical AI literacy that is engaging and inviting even as it includes critical perspectives. While the marketing of AI, which can also be seen as AI hype, often emphasizes power and pleasure, and critique of AI is often seen as anti-power or anti-pleasure, this form of critical literacy involves a sense of power and pleasure for those engaged in the discourse - this is similar to Martin Weller's (2022) reasoning for using metaphors as a playful approach to understand new domains in edtech while resisting or subverting dominant marketing discourses. This can play an important role pedagogically as we seek to create an environment for classroom discourse that is invitational, enjoyable and open-ended but also interrogates power structures.

We saw how this ethos of play and exploration helped us build our own critical AI literacy both in regard to technical understanding of the nature and limitations of language models and our sense of the range of uses and relationships to AI that are possible. The second question we asked ourselves about each metaphor was "Where does it help us understand AI?" Some responses, such as "stochastic parrot" help us understand the underlying statistical and random nature of how LLMs were built and how they work, which can also help us achieve a more critical understanding of why their outputs cannot be 100% credible or trustworthy (Bender et al, 2021; Hannigan, et al., forthcoming). The calculator metaphor, while unhelpful in its simplicity, can be beneficial in educational circles to discuss what "threshold" or "foundational" learning is needed before we can "permit" learners to start using a tool after they've understood the fundamentals needed in order to use it appropriately. Metaphors such as "wolf in sheep's clothing" help us question platforms that appear benign on the outside, but may have sinister or harmful processes that are hidden from us as end users. The "colonizing loudspeaker" metaphor reminds us of the harmful impact of technology platforms that reproduce white Western epistemologies and language while they "drown out the language of missing people who do not have the global capital to increase the volume of their utterances"(Owusu-Ansah, 2023).

What kind of literacies do these metaphors encourage in the classroom? How does each metaphor nudge our teaching and students' engagement with AI? The questions we asked about each metaphor point toward technical literacies as well as the ethical literacies described above under theme 2. Our explorations of metaphor brought us to a multidimensional discussion of the technical aspects of LLMs. Interestingly, we often chose different technical features to highlight in our reactions to the same metaphors. For example, about Ted Chiang (2023) metaphor of the blurry JPEG of the web, Author 3 wrote "How errors arise because LLMs approximate without understanding" and Author 1 wrote "that GPT is a lossy compression of language on the web; that it distorts and is thus imperfect; it hallucinates; it also highlights its statistical qualities." Also, the metaphors highlighted different kinds of technical questions. For example, the western museum metaphor focused our attention on the origins of the data used to train LLMs whereas the shoggoth with a smiley face metaphor was a way to describe RLHF's relation to the largely unknown structure of a pretrained LLM. The cake analogy (Bali, 2023) helps learners and educators question what processes and products are the essential goals of a learning experience, and which elements may be comfortably and safely relegated to AI; it also highlights the diversity of pedagogical contexts and needs, and why it is so difficult to come up with institution-wide guidelines around AI use.

Since we invited examination of emotional connotations and associative responses to the metaphors as well as reflecting on our own positionality in our writing process, our discussion built on our situated, individual responses to analyze power, positionality, and the practical implications of LLMs and AI for marginalized groups. For example, Author 1 reflected on the "leveling the playing field" metaphor in a way that took us away from a more generic discussion of power (of the kind Author 3 put in her responses) into a greater sense of urgency and concreteness about what is at stake. Author 1's response engaged Trimbur's "Literacy and the discourse of crisis" (1991) which showed that "moments of literacy crisis hide cultural anxieties of the middle class about their potential downward mobility and the potential upward mobility of other classes. With LLMs we are seeing the potential automation of essayist literacy which has historically been the rise to power for the middle classes and coding literacy, which more recently has been the source of power for a techno-cratic middle class." Exploring his own class anxiety and his concern as a teacher for multilingual students' position facing discrimination and seeking to use technology as a way to counter it allowed a way to bring to life a common theoretical concern in the idea of critical literacy. This kind of participatory, positionally grounded response to metaphor seemed like it could contribute to a much deeper visceral understanding of the mixed power implications of LLMs that could be part of a more meaningful critical AI literacy.

Thus I (Author 1) must be honest and note that my desire for critical conversations on LLMs are also perhaps informed by these class anxieties as I see literacies I have acquired with a lot of hard work, privilege and which have given me privilege, potentially becoming redundant. This is why I was especially drawn to the metaphor of "leveling the playing field" which had emerged based on my experiences with several multilingual students in my classroom. For some of these students, who have faced discrimination in grading because of their lack of familiarity with Standard American English (S.A.E), they found the use of ChatGPT like technologies to be especially liberating because in some ways, these "leveled the playing field" for them and allowed their ideas to be expressed in ways that would be taken more seriously inside American classrooms.

As we develop critical AI literacies, we should be open to the ways in which digital technologies have historically been repurposed by marginalized populations for social mobility as well as social justice. Social media for example, is both a tool for capitalist surveillance, but has also spurred protests against authoritarian regimes and encouraged critical dialogue across a range of political positions. We are interested in metaphors that reveal harm but also ways in which negative metaphors miss good things, etc.

Here we note as well the way in which critical AI literacies intersect with and complement data literacies. In their analysis of drivers and barriers that impact the implementation of ed-tech products based on data analytics and machine learning in educational institutions, Renz and Hilbig (2020) had found that a major barrier to implementation is a "lack of data understanding and insufficient data sovereignty" (p.17) in potential users. Through such exercises involving metaphor, teachers can help bridge students' existing knowledge of other domains to build understanding of how contemporary digital AI tools use and produce data. The critical dialogue around metaphor in turn can enhance students' ability to agentively engage with these technologies.

**CONCLUSION, IMPLICATIONS AND SUGGESTIONS**

Our explorations point to the value of metaphor as a stimulus to discussion of AI. The kind of ongoing dialogue with metaphor we have modeled is playful, critical, nuanced, and reflective of positionality. The need to imaginatively understand AI will continue, and the temptations and pressures to adopt misleading metaphors will also continue. The practice of interrogating and evolving our understanding through discussion of metaphors can be foundational for critical AI literacy.

A playful, curious, and open approach to exploring metaphors' implications we describe is not just valuable for its responsiveness to positionality and to the rapidly changing qualities of AI. It is also valuable because it is appealing and inviting. This is important pedagogically because the affective domain has powerful effects on learning that are sometimes neglected (Pierre & Oughton, 2007). Wang et al. (2023) showed that creating a supportive environment is essential to enabling students' learning about AI. In their work, supportive environments consist of "facilitating conditions" like accessibility to technology, and "supportive social norms" which refers to the extent to which students perceive their social mentors' desire to want them to learn about AI. Our work helps extend this conversation on creating supportive environments for AI learning because the proposed metaphors are often accessible, intriguing, fun, funny, and surprising. The open-ended nature of the exercise may make it less stressful and pressured, and point toward the creative opportunity to brainstorm new metaphors.

Our approach also provides accessible ways to cultivate critical data literacy in students, responding to the concern we have noted from researchers about the limits of existing functional approaches to data and AI literacy (Raffaghelli et al., 2020); Zawacki-Rickter et al.,2019). Our approach continues to culminate literacy around functionality and technical structure of AI systems even as it invites critical analysis of the context and impacts of AI.

For educators interested in using metaphor in class, students can share, discuss, and/or read about some metaphors and then challenge students on their own to say what each metaphor misses and to come up with an additional metaphor or metaphors that addresses this gap or misunderstanding.  To use metaphor to discuss AI specifically, a good starting point is to collect (anonymously using a polling tool or in a Google doc, or orally, or in writing as an assignment) metaphors for AI that come to students immediately, before you have too much of a discussion on AI. This gives a baseline of where their thoughts are, and students can also be asked if this is a metaphor they've heard/read, or one they've come up with spontaneously. One can discuss the kinds of impressions each of these metaphors have and how they may be both expressing how we feel about AI, or how saying them may shape how we feel about AI. Educators can also invite students to create their own AI meme featuring a metaphor of their choice and then write a rhetorical analysis of it that draws on research.

Educators can then use some of the metaphors shared here and discuss with students what we learn from each of the metaphors and these can be used throughout the semester where critical AI literacy is being developed. Learners can create a gameboard in the shape of the grid in our results section (or create their own grid with different dimensions on the X- and Y- axes) and have students in groups place notecards with metaphors on them along the dimensions of functional, critical and rhetorical as well as human, animal or partly human, and nonhuman. Then ask them to discuss and report back on any uncertainty or debates about where to place particular metaphors. Emphasizing dialogue over convergence is important here.

Towards the end of the semester, the educator may ask students to reflect again on which metaphors for AI resonated the most with them and whether they have new ones. This can be an in-class discussion, or a personal reflection. Learners may even be encouraged to compare two or more metaphors, where they lie on a matrix or continuum, and what the pros and cons are of using such metaphors in discourse. Write up scenarios where a person (with positionality described) needs to decide whether or how to use AI, perhaps specifically text generation. Then ask students in groups to make a case for which metaphor would be most helpful to that person in that situation as they try to find the right way to think about AI in that context. Learners may also be encouraged to role-play particular workplace or social situations where they are explaining AI or discussing it with others, and which metaphors might be useful for discussing it with, say, a marketing manager versus a six-year-old child. It is also important to discuss the limitations of any metaphor, and of metaphor a as learning strategy, so that we never stop at just the metaphor, but recognize the boundaries of each metaphor on its own. We hope to see future research into pedagogical approaches to critical AI literacy that combine functional and critical approaches. In addition, we think our approach may also be helpful starting point for researchers of public discourse of AI who want to design systematic studies to understand how public perceptions of AI, LLMs, and ChatGPT, are evolving in different communities across the globe over time and how that is shaping AI adoption and use across professional sectors.

## About the Author(s)

- Corresponding Author: Anuj, Gupta; University of Arizona, USA, anujgupta@arizona.edu, ORCID number: 0000-0002-5336-927X

- Yasser, Atef; American University in Cairo, Egypt, yasser.tamer@aucegypt.edu, ORCID number: 0000-0002-2755-8800

- Anna Mills; Cañada College, USA, millsanna@smccd.edu, ORCID number: 0009-0001-5094-1461

- Maha Bali; American University in Cairo, Egypt, bali@aucegypt.edu, ORCID number: 0000-0002-8142-7262

**Author's Contributions ([CRediT](CRediT))**
Anuj Gupta: project administration, conceptualization, data curation, formal analysis, investigation, methodology, writing—original draft preparation, writing—review and editing. Yasser Atef: conceptualization, data curation, formal analysis, investigation, methodology, writing—original draft preparation, writing—review and editing . Anna Mills: conceptualization, data curation, formal analysis, investigation, methodology, writing—original draft preparation, writing—review and editing. Maha Bali: conceptualization, data curation, formal analysis, investigation, methodology, writing—original draft preparation, writing—review and editing. All authors have read and agreed to the published version of the manuscript.

**Data Accessibility Statement**

The datasets generated and/or analysed during the current study are available in the
"Dataset 1: Metaphors of AI and their sources" repository: https://zenodo.org/records/10468056

**Ethics and Consent**
Since this analysis was auto-ethnographic in nature, we did not need to seek formal consent from anyone.

**Acknowledgements**
The authors thank various members of their social networks on X (Twitter) for participating in discussions about metaphors in particular and AI in general. They acknowledge their contributions, believing that the metaphor reflections were stimulated from these discussions. They also thank the people who attended their workshops on AI, as they shared insights that influenced in some way the reflections shared within this manuscript.

**Author(s) Notes**
Sections [abstract only] of this paper were generated with the assistance of OpenAI's GPT-4 and later heavilty edited by human authors.

**Funding Information**
Not applicable.

**Competing Interests**
The authors have no competing interests to declare